\documentclass[12pt,preprint]{aastex}
\usepackage{natbib}
\usepackage{graphicx}
\usepackage{amsmath}
\usepackage{rotating}
\usepackage{epsfig}
\def \dcop {\ensuremath {{\rm DCO^{+} }}}
\def \hthcop {\ensuremath{{\rm H^{13}CO^+}}}
\def \thco {\ensuremath{{\rm ^{13}CO}}}
\def \cdo  {\ensuremath {{\rm C^{18}O}}}

\begin{document}
\title{Supernova enhanced cosmic ray ionization and induced
  chemistry in a molecular cloud of W51C}
\author{ C. Ceccarelli$^1$, P. Hily-Blant$^1$, T.Montmerle$^{1,2}$,
  G.Dubus$^1$, Y.Gallant$^3$, A.Fiasson$^4$}
\altaffiltext{1}{UJF-Grenoble 1 / CNRS-INSU, Institut de
  Plan\'etologie et d'Astrophysique de Grenoble (IPAG) UMR 5274, 
  Grenoble, France \\
  Cecilia.Ceccarelli@obs.ujf-grenoble.fr, Pierre.Hily-Blant@obs.ujf-grenoble.fr,
  Guillaume.Dubus@obs.ujf-grenoble.fr}
\altaffiltext{2}{Institut d'Astrophysique de Paris, CNRS, France\\
   montmerl@iap.fr}
\altaffiltext{3}{Laboratoire de Physique Th\'eorique et
  Astroparticules, UMR 5207, CNRS/IN2P3, Université Montpellier II,
  France\\
  Yves.GALLANT@lpta.in2p3.fr}
\altaffiltext{4}{LAPP, Laboratoire d'Annecy-le-Vieux de Physique des
  Particules, UMR/IN2P3-CNRS, Universite de Savoie, 
  Annecy-le-Vieux, France\\
  fiasson@lapp.in2p3.fr}
    \date{Received - ; accepted -}
\begin{abstract}
  Cosmic rays pervade the Galaxy and are thought to be accelerated in
  supernova shocks. The interaction of cosmic rays with dense
  interstellar matter has two important effects: 1) high energy
  ($\gtrsim$ 1 GeV) protons produce $\gamma$-rays by $\pi^0$-meson
  decay; 2) low energy ($\lesssim$ 1 GeV) cosmic rays (protons and
  electrons) ionize the gas.  We present here new observations towards
  a molecular cloud close to the W51C supernova remnant and associated
  with a recently discovered TeV $\gamma$-ray source.  Our
  observations show that the cloud ionization degree is highly
  enhanced, implying a cosmic ray ionization rate $\sim 10^{-15}$
  s$^{-1}$, i.e. 100 times larger than the standard value in molecular
  clouds. This is consistent with the idea that the cloud is
  irradiated by an enhanced flux of freshly accelerated low-energy
  cosmic rays. In addition, the observed high cosmic ray ionization
  rate leads to an instability in the chemistry of the cloud, which
  keeps the electron fraction high, $\sim 10^{-5}$, in a large
  fraction (Av$\geq 6$mag) of the cloud and low, $\sim 10^{-7}$, in
  the interior. The two states have been predicted in the literature
  as high- and low-ionization phases (HIP and LIP). This is the
  observational evidence of their simultaneous presence in a cloud.
 \end{abstract}

%

\section{Introduction}\label{sec:introduction}

Cosmic rays (CR) pervade the Galaxy.  In the dense interstellar medium
(ISM), they play a crucial role as they are the primary source of
ionization, starting a complex chemistry which leads to hundreds
molecules, and influencing the star and planet formation processes by
creating ions, which in turn are coupled with the magnetic fields,
this regulating gravitational collapse.

There is little doubt that CR are accelerated in the expanding shocks
of supernova remnants (SNRs, e.g. Hillas 2005; Caprioli et
al. 2011). High energy ($\ga$ 1 GeV) CR (mainly protons) interact with
hydrogen atoms and produce $\gamma$ rays via $\pi^0$ decay (Hayakawa
1952; Stecker 1971). In this case, the predicted $\gamma$-ray
luminosity L$_\gamma$ is proportional to the $\pi^0$-decay
$\gamma$-ray emissivity, which depends on the local CR density, and to
the irradiated molecular cloud mass (e.g. Aharonian \& Atoyan
1996). Hence, giant molecular clouds close to, or even better,
penetrated by SNRs can be bright $\gamma$-ray sources. Conversely,
molecular clouds associated with bright $\gamma$-ray sources can probe
enhanced CR densities. Several CR interaction sites have now been
tentatively identifed with the latest generation of GeV/TeV
observatories (e.g. Montmerle 2010). The main difficulty remains to
distinguish between emission from proton and/or bremsstrahlung or
inverse Compton from electrons. Even in well-documented cases such as
IC443 (Albert et al. 2007) and W28 (Aharonian et al. 2008) SNRs, the
GeV-TeV emission mechanism remains unclear. In other cases (in
particular for W51C; Abdo et al. 2009; Feinstein et al. 2009)
$\pi^0$-decay appears to be the dominant $\gamma$-ray emission
mechanism. In these cases, the derived local relativistic proton
density is very high, typically one to two orders of magnitude higher
than the average galactic CR density. Such high densities should also
have visible effects at lower CR energies ($\la$ 1 GeV), in the regime
where CR ionize molecular clouds (e.g. Kamae et al. 2006; Gabici et
al. 2009; Padovani et al. 2009; Fatuzzo et al. 2010).

We propose that the physical interaction between SN-accelerated
energetic particles and molecular gas, where an association is
suggested by the presence of a $\gamma$-ray source, can be
demonstrated by using the impact of low-energy CR on the cloud
chemistry.
The method is based on the determination of the ionization degree of
the molecular cloud, which in turn gives a measure of the CR
ionization rate $\zeta$ (to be compared with the ``standard'' value
$\zeta_0 \sim 10^{-17}$ s$^{-1}$ for dense clouds; Glassgold \& Langer
1974).  In dense gas, the ionization can be obtained by measurements
of the DCO$^+$/HCO$^+$ abundance ratio (e.g. Gu\'{e}lin et
al. 1977). Briefly, HCO$^+$ and DCO$^+$ are formed by the reaction of
CO with H$_3^+$ and H$_2$D$^+$ respectively: CO + H$_3^+$
$\rightarrow$ HCO$^+$ + H$_2$ and CO + H$_2$D$^+$ $\rightarrow$
DCO$^+$ + H$_2$. Consequently, the DCO$^+$/HCO$^+$ ratio directly
depends on the H$_2$D$^+$/H$_3^+$ only.  In molecular clouds, CR
ionize the gas by ionizing H and H$_2$ at a rate $\zeta$ and forming
the molecular ion H$_3^+$, whereas H$_2$D$^+$ is formed by the
reaction of H$_3^+$ with HD.  Since both molecules are destroyed by
the reaction with the most abundant neutral species, CO, and by the
recombination with electrons, the DCO$^+$/HCO$^+$ ratio is an almost
direct measure of the gas ionization degree.  The method has been
extensively applied to derive the ionization degree of molecular
clouds and dense cores. The measured values range between
$1\times10^{-8}$ and $1\times10^{-6}$, depending on the gas density,
leading to estimates of $\zeta$ between $10^{-18}$ and $10^{-16}$
s$^{-1}$ (Caselli et al. 1998: hereinafter CWTH98).  Finally, Indriolo
et al (2010) recently reported line absorption observations of
H$_3^+$, measuring the ionization in the {\it diffuse gas} of the
outer parts of the cloud associated with IC443. Their measurements
indicate values of $\zeta$ only five times larger than the average in
diffuse clouds. In this Letter, we report observations of the
DCO$^+$/HCO$^+$ ratio towards the {\it dense gas} of a cloud
interacting with the SNR W51C.

\section{W51C, a SNR interacting with a molecular cloud}\label{sec:selected-sources}

W51C (also known as J1923+141 and G49.2-0.7 in the literature) is a
well known SNR at a distance of $\sim 6$ kpc (Kundu \& Velusamy
1967). The SNR extends about 30$'$ (equivalent to $\sim$60 pc), is
about $3\times10^4$ yrs old (Koo et al. 1995) and is associated with a
molecular cloud, engulfed by the blast wave, whose mass is $10^4$
M$_\odot$ and average density $\sim10$ cm$^{-3}$ (Koo \& Moon 1997).
Five HII regions lie at the border of the SNR, the W51A and B star
forming regions. Observations of HI and CO line emission have shown
the presence of shocked material north-west, at about a distance of 10
pc from the center of the SNR (Koo et al. 1995; 1997), where OH maser
emission is also detected (Hewitt et al. 2008). In this work, we
targeted five positions of the molecular cloud (Table
  \ref{tab:obs-tel}). Four of them lie a few arcmin away from the
  shocked region, while one (point A) lies at the northern edge of it.
An extended ($\sim 28$ pc) GeV--TeV source has been detected by HESS
(Feinstein et al. 2009) and Fermi-LAT (Abdo et al. 2009), with a total
$\gamma$-ray luminosity of $\sim10^{36}$ erg s${-1}$cm$^{-2}$ (Abdo et
al. 2009), making W51C one of the most luminous $\gamma$-ray sources
of our Galaxy. Based on the observed GeV--TeV $\gamma$-ray spectrum,
Abdo et al. (2010) claim that almost 99$\%$ of the observed
$\gamma$-rays is due to the decay of $\pi^0$ mesons produced in
inelastic collisions between accelerated protons and target gas.

\section{Observations and results}\label{sec:observations-results}

We observed five positions roughly sampling the cloud overlapping the
HESS emission (\S 2; Tab. \ref{tab:obs-tel}).  To constrain the gas
physical conditions, we observed the $^{13}$CO and C$^{18}$O 1--0 and
2--1 transitions, while to measure the ionization degree, we observed
the H$^{13}$CO$^+$ 1--0 and DCO$^+$ 2--1 transitions.
The observations were obtained using the APEX and IRAM-30m telescopes
(Tab. \ref{tab:obs-tel}).  The amplitude calibration was done
typically every 15 min, and pointing and focus were checked every 1
and 2 hrs respectively ensuring $\sim 1-3''$ accuracy. 
The APEX spectra were obtained in the classical ON-OFF mode, where the
OFF at EQ 2000 coordinates 19:22:44.3, 14:05:50.0 shows no signal. The
Fast Fourier Transform Spectrometer facility was used as a backend.
For the IRAM observations, we used the VESPA autocorrelator, and the
frequency-switching mode.
All spectra were reduced using the CLASS package (Hily-Blant et al
2005) of the GILDAS software.
Residual bandpass effects were subtracted using low-order ($\leq$ 3)
polynomials. 
%
\begin{sidewaystable}[tb]
\footnotesize
\caption{\label{tab:obs-tel}Upper half table: Parameters of the
  observations.  Lower half table: Observed positions with coordinates
  in J2000 and B1950 and line intensities main-beam temperatures (in
  K). Number in parenthesis are statistical uncertainties at the
  1$\sigma$ level. Lines in the position C show two separate velocity
  components.  Notes:$^a$APEX observations; $^b$IRAM observations:
  note that the second velocity component of the point C is not
  detected because the line is at the border of the filter.  }
\begin{tabular}{@{}llcccccccc}
    \tableline
    \tableline
    Species & Transition & Frequency & Main beam   & HPBW  & Telescope & Journal & Res. & T$_{\rm sys}$ &\\
                &                   &  (GHz)       &  efficiency    & (arcsec) &  & & (MHz) & (K) & \\
    \tableline
    \cdo    & 1--0 & 109.782 & 0.79  & 22 & IRAM & May 2009 & 0.08 & $\sim 100$ &\\
    \cdo    & 2--1 & 219.560 & 0.75  & 28 & APEX & Dec 2008  & 0.5   & $\sim230$ &\\
    \thco   & 1--0 & 110.201 & 0.79  & 22 & IRAM & May 2009 & 0.08 & $\sim 100$ &\\
    \thco   & 2--1 & 220.399 & 0.55  & 12 & IRAM & May 2009 & 0.08 & $\sim 100$ &\\
    \thco   & 2--1 & 220.399 & 0.75  & 28 & APEX & Dec 2008  & 0.5   & $\sim230$ &\\
    \hthcop & 1--0 &  86.754 & 0.79  & 28 & IRAM & May 2009 & 0.08 & $\sim 100$ &\\
    \dcop   & 2--1 & 144.077 & 0.79  & 17 & IRAM & May 2009 & 0.08 & $\sim 100$ &\\ 
    \tableline\tableline
    Position &   Coordinates (J2000) & $v_{\rm LSR}$ & \multicolumn{3}{c}{$^{13}$CO} & 
    \multicolumn{2}{c}{C$^{18}$O} & H$^{13}$CO$^+$ & DCO$^+$ \\
      &  ~~~~~~~~~~~~~~~~(B1950) & & 1--0 & 2--1$^a$ & 2--1$^b$ & 1--0 & 2--1 & 1--0 & 2--1 \\
      \tableline
       A     & 19:22:53.8 ~+14:15:44 & 71.2 &  4.30(0.03) &  6.20(0.05) & 9.00(0.1) & 0.30(0.03) & 0.45(0.04) & 0.040(0.006) & --   (0.01) \\
             & 19:20:35.9 ~+14:09:54 & & & & & & & & \\
      B     & 19:22:31.0 ~+14:15:45 & 69.0 &  7.60(0.04) &  9.20(0.05) & 17.85(0.2) & 0.50(0.05) & 1.10(0.04) & --    (0.01)  & --   (0.02) \\
             & 19:20:13.1 ~+14:09:57 & & & & & & & & \\
      C1   &19:22:21.3 ~+14:05:12 & 69.0 & 14.00(0.04) & 19.40(0.10) & 18.80(0.1) & 2.00(0.03) & 4.00(0.06) & 0.23(0.01)  & --   (0.02) \\
      C2  & 19:22:21.3 ~+14:05:12 & 73.0 &  7.70(0.04) & 14.80(0.10) & -               & 1.00(0.03) & 2.70(0.06) & 0.08(0.01)  & --   (0.02) \\
             & 19:20:03.2 ~+13:59:24 & & & & & & & & \\
      D    & 19:23:27.3 ~+14:16:24 & 65.5 &  2.80(0.07) &  4.20(0.07) & 6.30(0.1) & 0.15(0.02) & 0.30(0.06) & --   (0.007)  & --   (0.01) \\
             & 19:21:09.4 ~+14:10:32 & & & & & & & & \\
      E     & 19:23:08.0 ~+14:20:00 & 67.0 & 11.70(0.10) & 15.00(0.10) & 15.6(0.1) & 2.20(0.06) & 4.20(0.06) & 0.60(0.01)   & 0.040(0.006)\\
             & 19:20:50.1 ~+14:14:09 & & & & & & & & \\
      \tableline
  \end{tabular}
\end{sidewaystable}
Table \ref{tab:obs-tel} summarises the observed signals at each point
and each observed transition.
%
\begin{table*}
  \caption{\label{tab:signals}Analysis results. [$^{13}$CO]/[C$^{18}$O]
    ratio, temperature, density, C$^{18}$O and A$_{\rm v}$
    (assuming [C$^{18}$O]/[H$_2$]=$2\times10^{-7}$ (Frerking et
    al. 1982) and the usual N(H$_2$)=A$_{\rm v} \times 0.95 \times
    10^{21}$ cm$^{-2}$ mag$^{-1}$), extent of the emitting region in
    arcsec, H$^{13}$CO$^+$ column density, and
    [DCO$^+$]/[H$^{13}$CO$^+$] abundance ratio (assuming
    [HCO$^+$]/[H$^{13}$CO$^+$]=50) as derived by the analysis of the
    observed lines described in the text. The intervals account for the
    statistical uncertainty on the signals and on the physical
    parameters.   To account for the
    uncertainty on the CO-to-Av ratio, the Av interval is increased by
    an additional 30\%. Note that we run models with temperatures between 10
    and 50 K and densities between 10$^3$ and 10$^5$ cm$^{-3}$ so that
    the lower limits applies to these intervals only.  }
\begin{tabular}{@{}ccccccccc}
   \tableline
     & [$^{13}$CO]/ & Temp & Density & N(C$^{18}$O) & Av & extent & N(H$^{13}$CO$^+$) & [DCO$^+$]/\\
              & [C$^{18}$O]   &      (K)           & ($10^3$cm$^{-3}$) & ($10^{15}$cm$^{-2}$) & (mag) & (arcsec) & ($10^{11}$cm$^{-2}$) &[HCO$^+$]\\ \hline
A   & 15 & $\geq10$ & $\geq1$ & 0.4--0.6& 2--6    & 19            & 3--4       & $\leq 0.06$\\
B   & 15 & $\geq32$ & $\geq40$& 1.5--2.5& 7--14  & 27            &$\leq 1$& - \\
C1 & 15 & 23--28 & 6--20     & 3.9--4.1 & 16--24  & $\geq$28 & 7--9       & $\leq0.002$\\
C2 & 10 & $\geq27$ & 4--15     & 1.8-2.5  & 8--14  & $\geq$28 & 6--8      & $\leq0.006$\\
D   & 10 & $\geq10$ & $\geq1$  & 0.2--0.9& 1--5    & 20            &$\leq 0.8$ & -\\
E    & 10 &  21--24    & 8--20     & 3.9--4.1 &16--24& $\geq$28 & 16--18  & 0.0012-0.0016\\
      \tableline
  \end{tabular}
\end{table*}
%

\begin{figure}[tbh]
  \centering
     \includegraphics[width=8cm,angle=0]{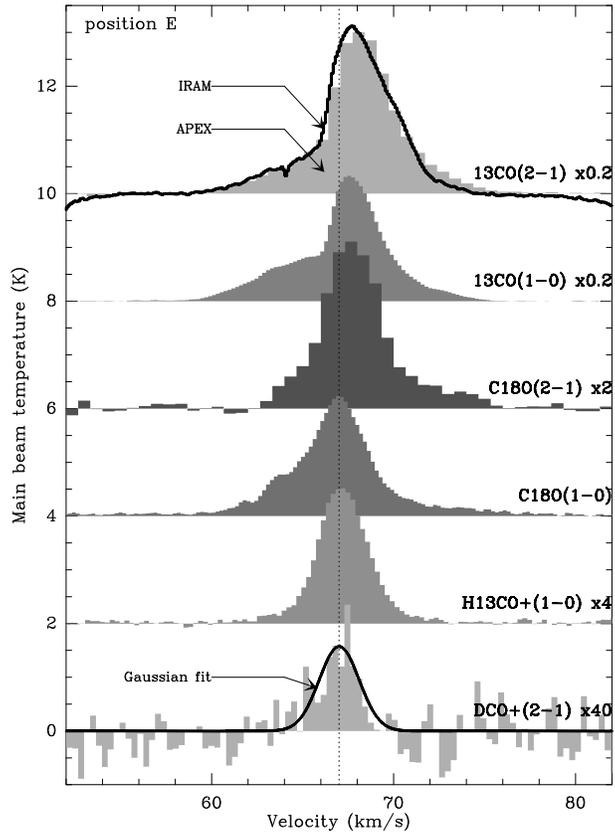}
     \caption{Observations towards position E. From top to bottom:
       $^{13}$CO 2--1, $^{13}$CO 1--0, C$^{18}$O 2--1, C$^{18}$O 1--0,
       H$^{13}$CO$^+$ 1--0 and DCO$^+$ 2--1. The signals are in main
       beam temperature K. The vertical line shows 67 km/s.}
  \label{fig:posE}
\end{figure}

CO emission is detected in the four observed transitions in all
points.  H$^{13}$CO$^+$ 1--0 emission is, on the contrary, detected
towards three positions only: A, C and E. Finally, DCO$^+$ 1--0
emission is detected only towards the position E. 
The spectra towards the position E are shown in Fig.\ref{fig:posE}.
In general, there are two clouds in the line of sight, at different
velocities v$_{\rm LRS}$ around 69 and around 73 km/s respectively.
The line widths are similar in all points and transitions, about 3
km/s. The comparison of the $^{13}$CO 2--1 observed with APEX and IRAM
gives a direct measurement of the emitting sizes. Points B, C and E
show extended ($\geq 28"$) emission, while in points A and D the sizes
are about 20$\arcsec$.
%

\section{Analysis}\label{sec:analysis}

\subsection{Physical conditions and column
  densities}\label{sec:physical-conditions}

We used the non-LTE LVG code described in Ceccarelli et al. (2003),
with the CO-H$_2$ collisional coefficients by Wernli et al. (2006) for
the first 6 CO levels and the temperature range 5-70 K. 
%
We run a grid of models with temperatures between 10 and 50 K,
densities between $10^3$ and 10$^5$ cm$^{-3}$, C$^{18}$O column
densities between $1\times10^{14}$ and $4\times10^{16}$ cm$^{-2}$ and
$^{13}$CO/C$^{18}$O ratio between 5 and 15.  For each point, we used
the extent derived by the comparison of the $^{13}$CO 2--1 line
intensity obtained at APEX and IRAM telescopes, respectively. The
values of the parameters minimising the $\chi^2$ are reported in
Tab. \ref{tab:signals} (the range corresponds to $\chi^2$=0.2
contours).
The H$^{13}$CO$^+$ column density has been derived by comparing the
observed signals with non-LTE LVG computations using the collisional
coefficients by Flower (1999) and the density and temperature derived
from the previous step. The H$^{12}$CO$^+$ column density is then
obtained by multiplying the N(H$^{13}$CO$^+$) by 50 (Koo \& Moon
1997). Note that the [DCO$^+$]/[HCO$^+$] ratio very little depends on
the actual gas temperature, for the range of temperatures considered
here.

The clouds in the direction of points C1 and E have temperatures
somewhat larger than 20 K, while the temperature is unconstrained in
the other points.  The densities are around $10^4$ cm$^{-3}$ in points
C1, C2 and E, and unconstrained in the other points. The visual
extinction A$_{\rm v}$ is few mag in points A and D, and it is larger than 10
mag in the other points. It is interesting and intriguing that points
C1 and E have similar physical conditions but different column
densities of HCO$^+$ and DCO$^+$.

\subsection{Ionization degree and CR Ionization rate}\label{sec:ioniz-degr-point}

Using the analytical method mentioned in \S 1 (Eqs. 1 and 2 in
CWTH98), from the N(DCO$^+$)/N(H$^{13}$CO$^+$) ratio\footnote{We used
  the rate coefficients reported in database KIDA ({\it
    http://kida.obs.u-bordeaux1.fr/}.}  we derive an extremely high
ionization degree toward the point E, $x(e)\gtrsim
2\times10^{-5}$. This value is one or two orders of magnitude larger
than those derived in other dark clouds by CWTH98. That point E is
``special'' with respect to the dark clouds of the CWTH sample is
indeed already clear from the observed signals: in those clouds the
DCO$^+$ 1-0 line intensity is of the same order of magnitude than the
H$^{13}$CO$^+$ 1-0 line intensity (Butner et al. 1995), whereas we
observed more than ten times lower DCO$^+$ 1-0 than the H$^{13}$CO$^+$
1-0 line intensity. This is a hallmark that the ionization degree is
much higher, and does not leave much of uncertainty on this
conclusion.  However, the analytical method is an oversimplification,
especially for high ionization degrees, as more routes of HCO$^+$
formation become important in addition to the H$_3^+$+CO one.

\begin{figure}[tbh]
  \centering
     \includegraphics[width=8.5cm]{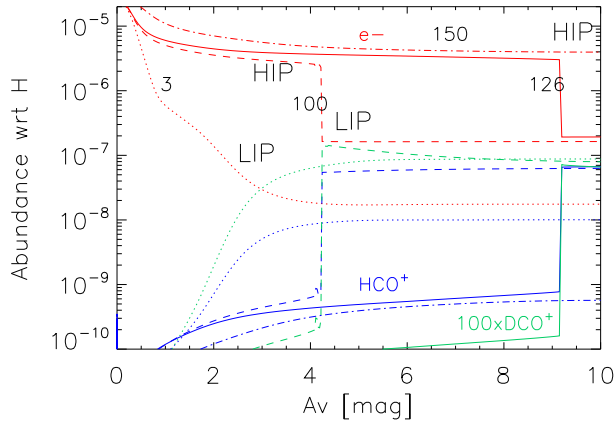}
     \caption{Structure of a 20 mag cloud of $1\times10^4$ cm$^{-3}$
       density illuminated by the IS UV field.  Electrons (red
       curves), HCO$^+$ (blue curves) and 100xDCO$^+$ (green curves)
       abundances (with respect to H nuclei) for four different values
       of the CR ionization rate: 3 (dotted line), 100 (dashed line),
       126 (solid line) and 150 (dashed-dotted line) $\times 10^{-17}$
       s$^{-1}$. The LIP and HIP regions in the four cases are also
       marked (see text). The computations refer to the metal low
       abundances case. Only the 10 mag from the edge of the cloud are
       shown. Note that the $\zeta$=126$\times 10^{-17}$ s$^{-1}$ case
       likely describes the situation towards point E.}
  \label{fig:family}
\end{figure}

In order to correctly evaluate the chemical structure of the cloud
towards the point E, including the penetration of the interstellar
(IS) UV photons and a (more) complete chemical network, we used the
photo-dissociation region (PDR) models by Le Petit et
al. (2006)\footnote{The code is available on the web site {\it
    http://pdr.obspm.fr/PDRcode.html}}.  To help facilitate the
discussion, Figure \ref{fig:family} shows the structure of a 20 mag
cloud whose density is $1\times10^4$ cm$^{-3}$, like point E,
illuminated on the two sides by the IS UV field \footnote{Higher
  values of the UV field would change the skin of the cloud only, so
  that we did not vary this parameter.} and with different CR
ionization rates. As found by several previous studies, for a low
value of $\zeta$ ($\lesssim 10^{-16}$ s$^{-1}$) and a low abundance of
metals, the ionization through the cloud is governed by the UV photons
penetration and depends on the abundance of carbon (A$_{\rm v}\leq$2
mag) and sulphur (2$\leq$A$_{\rm v}\leq$4 mag), and on the CRs
ionization deeper inside. The ionization at the interior of the cloud
is low, $10^{-8}$--$10^{-7}$, in a phase called LIP (low ionization
phase) in the literature (Pineau des For\^ets et al. 1992). This
situation is represented by the case $\zeta$=$3\times10^{-17}$
s$^{-1}$ in Fig. \ref{fig:family}.  In general, as $\zeta$ increases
the ionization in the cloud interior (A$_{\rm v} \geq 2$ mag)
increases roughly quadratically until it reaches a critical point
where it abruptly jumps to high values, in a phase called HIP (high
ionization phase). The LIP/HIP transition is a result of the
non-linear nature of the chemical network of molecular gas (e.g. Boger
\& Sternberg 2006; Rollig et al. 2007), which have been claimed to
present a bistable nature (Le Bourlot et al. 1993). The spatial
location of this transition depends upon various control parameters,
amongs which the density, the abundance of He and metals, and the CR
ionization rate (e.g. Wakelam et al. 2006). In practice, for a cloud
of a given density, the LIP/HIP jump is governed by the metal
abundances, specifically S, Si, Fe and Mg going inwards in the cloud,
as metals are responsible for the (non-linear) charge exchange. As
shown by Fig. \ref{fig:family}, the zones closer to the cloud edge
flip from LIP to HIP first, and the LIP/HIP jump moves further into
the cloud with increasing $\zeta$, until the whole cloud is in the
HIP, regardless the metal abundances.

The HCO$^+$ abundance is affected by $\zeta$ (the larger the $\zeta$
the larger the HCO$^+$ formation rate) and by the presence of the HIP
region (the larger the ionization degree the larger the HCO$^+$
destruction rate). In general, the HCO$^+$ abundance is roughly
divided into three zones: in the skin of the cloud, it is very low
because of the photo-dissociation by the UV photons, deeper in the
cloud it increases but the high ionization in the HIP region keeps the
HCO$^+$ abundance low, and at still larger Av it increases, when
eventually the gas jumps into the LIP.

\begin{figure}[tbh]
  \centering
     \includegraphics[width=8.5cm]{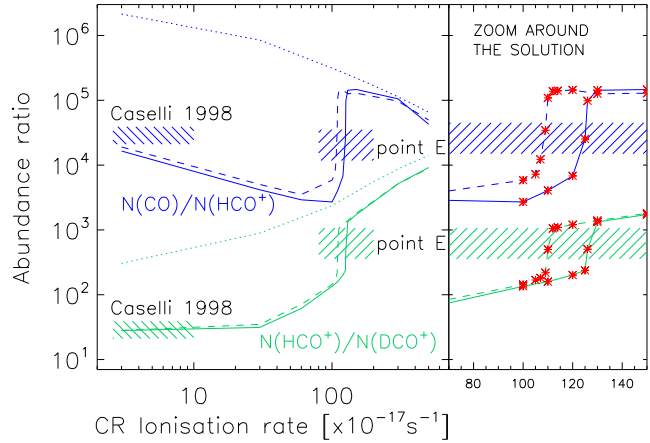}
     \caption{Observations against PDR model predictions:
       N(CO)/N(HCO$^+$ (blue) and N(HCO$^+$)/N(DCO$^+$) (green)
       integrated over the 20 mag cloud.  Three families of
       theoretical curves are reported with different metal
       abundances, following Wakelam et al (2006; see also CWTH98):
       high- (dotted), low- (solid) and intermediate- (dashed)
       abundances.  The boxes show the observed values towards the
       point E and, for comparison, in the CWTH98 sample. The right
       panel shows a zoom around the solution.  The red stars show
       the computed models.
}
     \label{fig:cr-constrain}
\end{figure}

The ratio N(CO)/N(HCO$^+$) of column densities, integrated over the 20
mag model cloud, is shown in Fig. \ref{fig:cr-constrain}. It decreases
with $\zeta$ (as the HCO$^+$ abundance increases), until a critical
point (at $\zeta\sim1.0-1.3\times10^{-15}$ s$^{-1}$ depending on the
metals abundance) where the whole cloud is in the HIP state and the
HCO$^+$ abundance becomes very low all across the whole cloud. Note
that for high metal abundances the cloud is always in the HIP
state. As noted by Wakelam et al. (2006), the ionization distribution
is bimodal with the metal abundances. In fact, the theoretical
N(CO)/N(HCO$^+$) ratio either follows the high or the intermediate
metal abundance curves of our Fig. \ref{fig:cr-constrain}.

The DCO$^+$ abundance follows qualitatively the behaviour of HCO$^+$
and the HCO$^+$/DCO$^+$ ratio across the cloud depends on the
  ionisation degree ($\sim 100$ in the LIP and $\geq 1000$ in the HIP
  zones, respectively).
The dependence of the N(HCO$^+$)/N(DCO$^+$) ratio as function of
$\zeta$ is shown in Fig. \ref{fig:cr-constrain}. In practice, the
  values of N(CO)/N(HCO$^+$) and N(HCO$^+$)/N(DCO$^+$) depend on the
  relative weight of the LIP-to-HIP zones in the integral over the
  cloud.

The comparison between the predicted N(CO)/N(HCO$^+$) and
N(HCO$^+$)/N(DCO$^+$) ratios with the values observed towards the
point E provides a value of $\zeta \sim$ 1.0--1.3
$\times 10^{-15}$ s$^{-1}$.
The likely structure of the cloud in the point E is shown in
Fig. \ref{fig:family}.  In practice, a large fraction of the cloud is
the HIP state whereas the rest is in the LIP state.  We stress,
  however, that, given the high non-linearity of the equations, the
  exact value of $\zeta$ which fits the observations depends on the
  details of the adopted model (Wakelam et al. 2006; Boger \&
  Sternberg 2006). Regardless of these details, the modeling indicates
  a situation where high (HIP) and low (LIP) ionisation must be both
  present, to give the observed HCO$^+$/DCO$^+$value which is an
  intermediate value of the two regions. The relative amounts of
  HIP/LIP may change with other control parameter values, but the
  result is that the line of sight we probe must contain both LIP and
  HIP. Finally, CR heat the cloud as they deposit energy so that
the relatively large value of the gas temperature (21--24 K) toward
point E (Tab. \ref{tab:signals}) also favours the high $\zeta$.

Finally, the lower limit to the N(HCO$^+$)/N(DCO$^+$) towards point C1
suggests a value for $\zeta$ larger than in point E.

\section{Conclusions}

Our observations lead to two important conclusions:\\
1) The cloud towards the point E has a CR ionization rate
$\zeta\sim 10^{-15}$ s$^{-1}$, i.e. enhanced by 
about a factor 100 with respect to the standard value (Glassgold \&
Dalgarno 1974). This is the first time that such an high ionization
degree is measured in a dense molecular cloud. The fact that this
cloud coincides with a TeV $\gamma$-rays source close to a SNR
conforts the idea that it is irradiated by an enhanced flux of freshly
formed low-energy CR. We have identified a location close to a CR
accelerator with both enhanced low-energy CR ionization and high CR
$\gamma$-ray emission, suggesting novel ways to study the acceleration
and the diffusion of CR close to SN shocks.\\
2) The high low-energy CR flux induces the simultaneous presence of a
region of high- and low- ionization, called HIP and LIP in the
literature. To our best knowledge, this is the first time that such a
situation has been observed and offers the possibility to study in
more detail the chemistry of the two states simultaneously. Therefore,
a more general implication of this work is that TeV sources identified
with SNRs interacting with molecular clouds are promising sites not
only to study CR acceleration, but also to study unusual chemical
conditions.

\begin{acknowledgements}
\end{acknowledgements}


\begin{thebibliography}{58}

\bibitem[]{} 
Abdo A.A., Ackermann M., Ajello M.et al. 2009, ApJ 706, L1
\bibitem[]{} 
Aharonian F.A. \& Atoyan A.M. 1996, A\&A 309, 917
\bibitem[]{} 
Aharonian F.A., Akhperjanian A.G., Bazer-Bachi A.R. et al. 2008, A\&A
481, 401
\bibitem[]{} 
Albert J., Aliu E., Anderhub H. et al. 2007, ApJ 664, L87
\bibitem[]{} 
Boger G.I. \& Sternberg A. 2006, ApJ 645, 314
\bibitem[]{} 
Butner H.M., Lada E.A., Loren R.B. 1995, ApJ 448, 207
\bibitem[]{} 
Caprioli D., Blasi P., Amato E. 2011, APh 34, 447
\bibitem[]{} 
Caselli P., Walmsley C.M., Terzieva R., Herbst E. 1998, ApJ 499, 234 (CWTH98)
\bibitem[]{} 
Ceccarelli C., Maret S., Tielens A.G.G.M., Castets A., Caux E. 2003, A\&A 410, 587
\bibitem[]{} 
Fatuzzo M., Melia F., Todd E., Adams F.C. 2010, ApJ 725, 515
\bibitem[]{} 
Feinstein et al. 2009, AIPC 1112 54F
\bibitem[]{} 
Flower D.R. 1999, MNRAS 305, 651
\bibitem[]{} 
Gabici S., Aharonian F.A., Casanova S. 2009, MNRAS 396, 1629
\bibitem[]{} 
Glassgold A.E. \& Langer W.D. 1974, ApJ 193, 73
\bibitem[]{} 
Gu\'elin M., Langer W.D., Snell R.L., Wootten H.A. 1977, ApJ 217, L165
\bibitem[]{} 
Hayakawa S. 1952, PThPh 8, 571
\bibitem[]{} 
Hewitt J.W., Yusef-Zadeh F., Wardle M. 2008, ApJ 683, 189
\bibitem[]{} 
A.M. Hillas 2005, J. Phys. G: Nucl. Part. Phys., 31
\bibitem[]{} 
Hily-Blant P., Pety J. \& Guilloteau S. 2005, CLASS Manual, http://www.iram-institute.org/medias/uploads/class-evol1.pdf
\bibitem[]{} 
Le Bourlot J., Pineau des For\^ets G., Roueff E., Schilke P. 1993, ApJ 416, L87
\bibitem[]{} 
Le Petit F., Nehmé C., Le Bourlot J., Roueff E. 2006, ApJS 164, 506
\bibitem[]{} 
Kamae T., Karlsson N., Mizuno T., Abe T., Koi T. 2006, ApJ 647, 692
\bibitem[]{} 
Koo B-C., Kim K-T., Seward F.D. 1995, ApJ 447, 211
\bibitem[]{} 
Koo B-C. \& Moon D-S. 1997, ApJ 485, 263
\bibitem[]{} 
Kundu M.R. \& Velusamy T. 1967, AnAp 30, 59
\bibitem[]{} 
Montmerle T. 2010, ASPC 422, 85
\bibitem[]{} 
Padovani M., Walmsley C.M., Tafalla M., Galli D., Müller H.S.P. 2009, A\&A 505, 1199
\bibitem[]{} 
Pineau des For\^ets G., Roueff E., Flower D.R. 1992, MNRAS 258, 45
\bibitem[]{} 
Rollig M., Abel N.P., Bell T., Bensch F., Black J. et al. 2007, A\&A 467, 187
\bibitem[]{} 
Stecker F.W. 1971, Nature 234, 28
\bibitem[]{} 
Wakelam V., Herbst E., Selsis F., Massacrier G. 2006, A\&A 459, 813
\bibitem[]{} 
Wernli M., Valiron P., Faure A., Wiesenfeld L., Jankowski P., Szalewicz K. 2006 A\&A 446, 367
\end{thebibliography}
\end{document}